# Exploring the composition of icy bodies at the fringes of the solar system with next generation space telescopes

**Thematic Areas:**    ⊠ Planetary Systems

**Principal Author:**
Name: Richard J. Cartwright
Institution: SETI Institute / NASA Ames Research Center
Email: rcartwright@seti.org

**Co-authors:** (names and institutions)
Bryan Holler                Space Telescope Science Institute
Susan Benecchi              Planetary Science Institute
Roser Juanola-Parramon      NASA Goddard Space Flight Center
Giada Arney                 NASA Goddard Space Flight Center
Aki Roberge                 NASA Goddard Space Flight Center
Heidi Hammel                Association of Universities for Research in Astronomy

**Co-signers:**
Tracy Becker, Chloe Beddingfield, Dale Cruikshank, Katherine de Kleer, Paul Estrada, Amanda Hendrix, Jason Hofgartner, Stefanie Milam, Marc Neveu, Tom Nordheim, Andy Rivkin, Megan Schwamb, Francesca Scipioni, Linda Spilker, Matthew Tiscareno, David Trilling, Anne Verbiscer, Michael Wong

## 1. Abstract and Motivation

Determining the distribution and spectral signature of volatile ices and organics exposed on icy body surfaces can provide crucial clues for deciphering how the outer solar system formed and evolved (*e.g.*, [1, 2]). Icy objects with primordial surfaces retain valuable information about the state and variety of constituents present in the early solar system. Icy objects with more processed surfaces demonstrate how the original mixture of constituents has evolved through time. Over the past few decades, ground- and space-based telescope observations have probed the compositions of a wide range of icy objects with primordial or processed surfaces, revealing the presence of numerous volatile ices and organic residues (*e.g.*, [3]).

Although these telescope observations have advanced our understanding of icy bodies beyond Saturn, the sensitivity and spatial resolution of collected datasets are limited by the large heliocentric distances of these far-flung objects. The Voyager 2 and New Horizons missions returned a wealth of data from their encounters with the Uranus, Neptune, and Pluto systems, but the flyby nature of these encounters limited the spatial coverage and temporal resolution of collected datasets. Repeat observations, occurring over decadal timescales, have been conducted for some of these objects, in particular the large Uranian moons [4-7], Triton (*e.g.*, [8, 9]) and Pluto (*e.g.*, [10, 11]). However, many of the other volatile-rich, icy bodies beyond Saturn have been minimally observed in both time and resolution. Consequently, our understanding of how volatiles migrate across the surfaces of icy bodies in response to seasonal change is still poorly developed. Furthermore, most observations have focused on the visible (VIS, ~0.4 – 0.7 µm) and near-infrared (NIR, ~0.7 – 2.5 µm), with fewer observations at longer NIR wavelengths (~2.5 – 5.0 µm) and in the far to near ultraviolet (UV, ~0.1 – 0.4 µm), which represents a critical wavelength region for investigating modification of ices and organics by UV photolysis and charged particle radiolysis (*e.g.*, [12-14]).

Thus, our understanding of icy bodies beyond Saturn is limited by the capabilities of available facilities, and key questions regarding their surface compositions remain to be explored. The ground-based Extremely Large Telescopes (ELTs) coming online in the 2020's will provide unprecedented angular resolution in the NIR, but Earth's atmosphere is opaque in the UV and will limit the capabilities of ELTs in the VIS and longer NIR wavelength regions as well. Next generation space telescopes (NGSTs) with greater sensitivity and angular resolution in the UV, VIS, and longer NIR are therefore needed to help unveil the surface compositions of icy bodies residing at the fringes of our solar system. In the following sections we describe some of the key questions that could be addressed using NGST observations of ice giant satellites, icy dwarf planets, and mid-sized Trans-Neptunian Objects (TNOs) [***Table 1***].

*Table 1: Icy bodies across the outer solar system*

| Solar System Zone (helio. distance) | Satellite Type | Semi-major Axis ($10^5$ km) | *Angular Diameter (arcsec.) | Confirmed Constituents | **Predicted Constituents |
|---|---|---|---|---|---|
| Uranian System (19.22 AU) | Regular (ring) | 0.498 - 0.977 | ~0.001 - 0.012 | ? | Org., H.S., $H_2O$ |
| | Regular (classical) | 1.29 - 5.84 | ~0.04 - 0.12 | $H_2O$, $CO_2$ | Org., H.S., $NH_3$-bearing species |
| | Irregular | 42.8 - 209.0 | ~0.001 - 0.01 | ? | Org. H.S. $H_2O$ |
| Neptunian System (30.11 AU) | Regular (inner) | 0.482 - 1.18 | ~0.001 - 0.02 | ? | Org., H.S., $H_2O$ |
| | Irregular (Triton) | 3.55 | ~0.13 | $H_2O$, $CO_2$, $CH_4$, CO, $N_2$ | $C_2H_6$, HCN |
| | Regular (Nereid) | 55.1 | ~0.02 | $H_2O$ | Org., H.S. |
| | Irregular | 166.1 - 493.0 | ~0.001 - 0.02 | ? | Org., H.S., $H_2O$ |
| Trans-Neptunian Objects (> 30 AU) | **Object Type** | **Helio. Dist. (AU)** | | ‡ | ‡ |
| | Resonant | ~39.3, ~47.8 | ~0.09† | $H_2O$, $CH_4$, CO, $N_2$, $C_2H_6$, $C_2H_4$ | Org., H.S., $CH_3OH$, $NH_3$-bearing species |
| | Classical Belt | ~42 - 48 | ~0.04† | | |
| | Scattered Disk | < 40 perihelion | ~0.03† | | |
| | Distant Detached | > 40 perihelion | ~0.02† | | |

*As seen from Earth-Sun L2 Lagrange point in 2040. ** Org. = organics, H.S. = hydrated silicates. ‡For large TNOs, across all object classifications. †Represents the angular diameters of the largest members of these groups.

## 2.  Satellites of the Ice Giants

Uranus is orbited by five large and tidally-locked "classical" moons, Miranda, Ariel, Umbriel, Titania, and Oberon, in order of increasing orbital radius, respectively. Interior to these large moons, thirteen small moons orbit within Uranus' ring system. Far beyond the classical satellite zone, nine irregular satellites orbit Uranus on highly inclined and eccentric orbits.

Ground-based observations determined that the classical moons have surface compositions dominated by a mixture of $H_2O$ ice and dark, spectrally neutral material (*e.g.*, [15-18]). Laboratory experiments indicate that the dark material is likely carbonaceous in origin, with a spectral signature similar to charcoal [19]. Overprinting this base composition of dirty $H_2O$ ice, $CO_2$ ice has been detected, primarily on the trailing hemispheres of the inner moons, Ariel and Umbriel [4-6]. The distribution of $CO_2$ ice on these moons is broadly consistent with radiolytic generation via irradiation of native $H_2O$ ice and C-rich material by magnetospherically-embedded charged particles [5, 6]. Spectrally red material has also been detected, primarily on the leading hemispheres of the outer moons, Titania and Oberon [7, 20-22]. The distribution of red material is broadly consistent with accumulation of dust from retrograde irregular satellites [7, 23], which are spectrally redder than the classical moons (*e.g.*, [24-28]).

Although previous work supports these hypotheses, many questions remain to be explored by NGSTs. $CO_2$ ice is present on the trailing hemispheres of these moons, but its distribution is poorly constrained due to the lack of spatially-resolved, NIR spectra. The influence of charged particle radiolysis on icy Galilean and Saturnian moons has been investigated using UV spectrometers onboard the Galileo and Cassini spacecrafts, respectively, along with the Hubble Space Telescope (HST) (*e.g.*, [29]). However, the small number of available UV spectra of the Uranian satellites [30-31] have low signal-to-noise (S/N) and are not spatially-resolved, limiting our ability to investigate the role of charged particle radiolysis on their surface compositions.

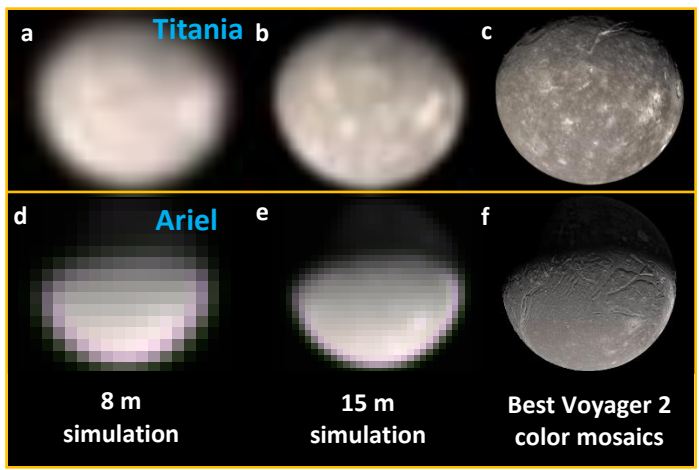

Red material was detected on these moons in images collected by the Imaging Science System (ISS) on Voyager 2 [20-22] [*Figure 1*], but their northern hemispheres were shrouded by winter darkness at the time of the Voyager 2 flyby, limiting our ability to constrain the global distribution of red material. Furthermore, Voyager 2 did not have a mapping spectrometer capable of collecting spatially-resolved spectra over visible and near-infrared wavelengths (~0.4 – 5.0 µm), limiting our ability to determine the composition of the red material. Based on the analysis of data collected by Cassini's Visible and Infrared Mapping Spectrometer (VIMS), spectrally red, organic material detected on icy Saturnian moons, (*e.g.*, [32-33]) displays weak absorption bands (due to C-H stretching in the ~3.3 – 3.5 µm region), but available facilities lack the sensitivity to reveal comparable abundances of organic residues on the more distant Uranian satellites.

*Figure 1: Resampled and real images of Titania (**a-c**) and Ariel (**d-f**). Real images (c, f) are best color Voyager 2/ISS image mosaics [78, 79]. The resampled images (a-b, d-e) simulate what these moons would look like as seen by a space telescope with an 8 m and 15 m aperture from the Earth-Sun L2 point.*

Far less is known about the compositions of Uranus' smaller ring moons and irregular satellites, which are too faint for high S/N spectroscopic observations with current facilities. The available VIS and NIR photometric datasets indicate that the ring moons are dark and neutrally sloped with slight reductions in albedo at 1.5 and 2.0 µm, hinting at the presence of weak $H_2O$ ice features [34]. The irregular satellites of Uranus are likely captured objects, either from nearby heliocentric sources (*e.g.*, [35]) or from the Kuiper Belt (*e.g.*, [36]). Photometric results indicate that Uranus' irregular moons are dark and red (e.g., [24-28]), with redder VIS/NIR colors than the irregular satellites of the other giant planets [37]. Low S/N spectra of the largest Uranian irregular Sycorax hint at the presence of $H_2O$ ice [38], but no spectra exist for the other, fainter moons.

Neptune is orbited by seven small, prograde inner moons, the close-in retrograde moon Triton, and prograde Nereid, which has a highly elliptical orbit. Like Uranus, Neptune also possesses a group of "normal" irregular satellites, which orbit at much greater distances than Triton and Nereid and are likely captured objects [*e.g.*, 35, 36]. The highly circularized, retrograde orbit of Triton has long been cited as strong evidence for its origin in the Kuiper Belt and subsequent capture by Neptune (*e.g.*, [39]). Furthermore, as revealed by Voyager 2, the non-spherical shapes and debris-strewn, asteroid-like surfaces of the inner Neptunian moons, and the large eccentricity of Nereid, could have resulted from catastrophic perturbation of Neptune's native satellite system during the Triton capture event. Aside from numerous observations of the large and bright Triton, the Neptunian satellite system is mostly unexplored, with only a handful of published studies that have investigated the compositions and processes modifying the surfaces of these moons. Because of their small sizes and proximity to Neptune, investigation of the inner moons' compositions is limited to broadband VIS and NIR spectrophotometry, primarily from HST datasets (*e.g.*, [40]). Similarly, investigation of the five small and distant irregular satellites' compositions has been limited to analysis of photometric datasets (*e.g.*, [26, 37, 41]). Some low S/N spectra of Nereid exist, indicating that a mixture of dark material and $H_2O$ ice is present [42], but more detailed spectral analyses are hampered by the low quality of these data.

Therefore, spatially-resolved, high S/N spectra spanning the UV, VIS, and NIR (~0.1 – 5.0 µm), represent crucial datasets for investigating the distribution and origin of volatile ices and organics on the satellites of Uranus and Neptune. The far superior PSFs provided by NGSTs (compared to current facilities) will significantly reduce scattered light from Uranus and Neptune, allowing for high S/N, spectroscopic observations of their inner moons. Similarly, high S/N spectral observations of these planets' irregular satellite populations would help determine their compositions, providing new clues regarding the origin of these likely captured objects.

## 3. Icy Dwarf Planets

Neptune's largest satellite, Triton, is thought to be a large TNO that was captured early in solar system history (*e.g.*, [39]). The Voyager 2 flyby revealed a surface of diverse terrains and geyser activity on short timescales (*e.g.*, [43]). More recently, an increase in volatile ices was observed over the period of a decade as the sub-solar latitude shifted northwards [9]. The diffraction limit for a 15-m NGST, results in ~140 km per resolution element, or about 300 pixels covering the surface area of Triton's disk [**Figure 2**]. Imaging observations made over long timescales can be used to track regional albedo changes across Triton's surface as the sub-solar latitude moves toward the equator over the next several decades, a viewing geometry never before observed in detail. Previous work has identified clear changes in regional VIS albedos between observations made by Voyager 2 in 1989 and HST in 2005, suggesting volatile transport on Triton [44].

Pluto was visited by the New Horizons spacecraft in July 2015. Although the Pluto flyby provided a wealth of spatially-resolved, high S/N images and spectra, no temporal changes were identified during the short timescale of the event [45]. With New Horizons well beyond Pluto and no follow-on missions planned, the only means to study temporal changes on Pluto is to use ground- and space-based facilities. High-resolution imaging and spectroscopy will provide the best means to quantify seasonal changes across Pluto. A 15-m telescope would provide ~185 km per resolution element with Pluto at 38 AU [*Figure 2*]. VIS albedo maps made with NGSTs could be compared to existing maps [46] to identify changes in regional albedos. It has been predicted that Pluto's northern hemisphere will be devoid of all volatile ices (*e.g.*, $CH_4$, CO, $N_2$) by 2030 [47], and a facility operating in the 2040's would be ideal for investigating this prediction.

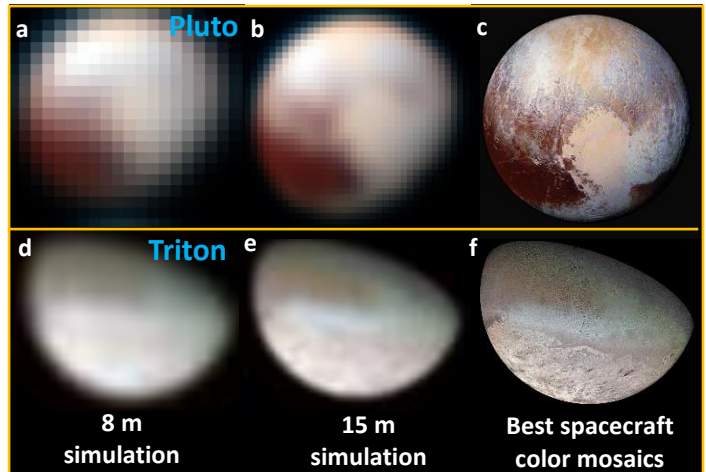

*Figure 2:* Resampled and real images of Pluto (***a-c***) and Triton (***d-f***). Real images are best color New Horizons [80, 81] (c) and Voyager 2 [78, 79] (f) image mosaics. The resampled images (a-b, d-e) simulate what these objects would look like as seen by a space telescope with an 8 m and 15 m aperture from the Earth-Sun L2 point.

Haumea is the only TNO known to have rings [48]. However, these rings are too faint for direct detection with available facilities and can only be studied through occultations. In 2040, Haumea will be at an observer-centric distance of 47 AU, so each resolution element of a 15-m telescope will cover ~230 km. Haumea's rings orbit with a semi-major axis of ~2300 km and are ~70 km thick [48], meaning that approximately 10 resolution elements would separate the rings from the center of Haumea. The rings contribute ~5% of the total brightness of Haumea and would be easily detectable in VIS imaging using a 15-m NGST.

Monitoring campaigns of VIS albedos can be used to track the seasonal migration of volatiles. At present, this type of study is only possible for Triton and Pluto with HST [44, 46]. A 15-m telescope will be able to conduct monitoring campaigns for more distant TNOs. For example, Makemake and Eris both have low, but non-zero light curve amplitudes (*e.g.*, [49, 50]). For Eris, a ~10% difference in albedo between the hemispheres representing the maximum and minimum in the light curve has been detected, pointing to a non-uniform distribution of volatile ices [51]. A similar situation can be assumed for Makemake. A 15-m NGST could obtain 30 and 20 pixels on the disks of Eris and Makemake, respectively, in the 2040's. Investigation of the spatial distribution of ices across the surfaces of these TNOs would contribute to our understanding of the timescale for volatile transport at larger heliocentric distances.

Sedna is the archetype of the dynamical class known as "sednoids," which have perihelia greater than 50 AU [52-54]. Sedna's aphelion is ~936 AU, meaning it spends most of its time beyond the heliopause, experiencing a radiation environment unique from closer-in TNOs. Ground-based spectroscopy of Sedna suggests that its surface is dominated by $CH_4$ and $N_2$ [55, 56], but the large heliocentric distance of Sedna and the other sednoids limits the S/N of collected datasets. NGSTs are needed to determine the compositions of these extremely distant objects.

## 4. Mid-Sized TNOs

Along with the large icy dwarf planets, the trans-Neptunian region includes a treasure box of mid-sized icy objects at the far reaches of our solar system. Since the discovery of the first TNO in 1992 [57], objects throughout this region of space have provided evidence for dynamical interaction with the giant planets [58-62] as well as signatures of both primitive surfaces [63] and more recent surface processing [64]. A wide range of volatile ices have been detected on mid-sized TNOs, and recent observations suggest that some of these objects have hydrated silicates exposed on their surfaces as well (*e.g.*, [65, 66]) [*Table 1*]. Therefore, the next steps for physical understanding of the processes operating on mid-sized and smaller TNOs will come from high S/N (≥ 50) spectra spanning a wide wavelength range (~0.1 – 5.0 µm), as well as low temperature (~50 K) laboratory studies of ices [67].

Dynamically, these objects are found in circular orbits (Classical), resonant orbits with Neptune (Resonant), highly inclined and elliptical orbits (Scattered and Detached), as well as giant planet crossing orbits (Centaur) [68-70]. At the limits of modern observational equipment, the objects we have directly imaged range in size from 27 km (2014 MU$_{69}$) to ~2300 km (Pluto and Eris).

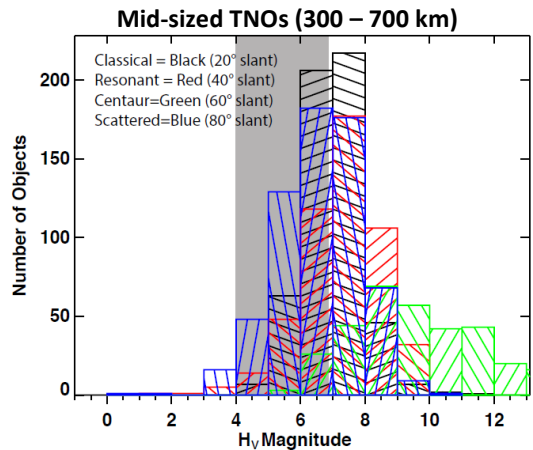

*Figure 3*: *Histogram of the numbers of TNOs by H$_V$ magnitude with respect to dynamical classification (Classical=black, Resonant=red, Centaur=green and Scattered=blue). Gray-toned region represents the mid-sized TNOs.*

Observations have primarily focused on the largest objects because they provide the highest S/N, allowing for the highest fidelity for physical interpretations. However, the largest objects (H ≥ 3.5, assuming an albedo of 0.1) are not necessarily the most representative of the population, rather they are end-members [*Figure 3*]. Surveys of more typically sized TNOs have mostly been in the optical [71], infrared [72], and thermal wavelengths [73]. Observations of mid-sized TNOs such as 19521 Chaos, 28978 Ixion, 229762 G!kúnǁ'hòmdímà, 20000 Varuna, 38628 Huya, and 79360 Sila-Nunam, to name a few, over a broader range of wavelengths would provide far more relevant results for extrapolation to smaller objects, as well as more fully sampling objects in varying dynamical regions of the trans-Neptunian region. If such observations are of objects for which physical properties can also be derived – *e.g.*, due to object binarity (88 TNO binaries are currently known) which allows us to estimate densities [74], or occultation measurements which can provide direct diameter and shape information [75], then we can build even greater confidence in our interpretation of physical processes at work on icy bodies in the far reaches of our solar system. Likewise, all TNOs exhibit rotational variations [76, 77], determined primarily over VIS wavelengths. However, there is a degeneracy in most of these observations between object shape and surface albedo distribution. Sometimes lightcurve shape can make breaking this degeneracy feasible, but many observations over a long baseline are required to do this effectively. Alternately, observations in two wavelength regimes (VIS and NIR) can also break this degeneracy.

## 5. Closing Remarks

In summary, to fully determine the compositions of icy bodies beyond Saturn, spatially-resolved, high S/N spectral datasets that span the UV, VIS, and NIR are needed. Ambitious projects that will develop large (8-m and 15-m class) NGSTs are required to collect these high-quality spectral datasets. With this spectral information in hand, we will be able to identify the distribution and spectral signature of volatile ices and organics on a wide swath of icy objects. Investigation of these different icy bodies will provide new insight into the origin and evolution of the constituents that comprised the early outer solar system.